\documentclass[aps,superscriptaddress,amsmath,amssymb,showpacs,twocolumn]{revtex4-1}
\usepackage{graphicx}

\usepackage[utf8]{inputenc}
\usepackage[T1]{fontenc}
\usepackage{mathtools}

\newcommand{\beq}{\begin{equation}}
\newcommand{\eeq}{\end{equation}}
\newcommand{\beqa}{\begin{eqnarray}}
\newcommand{\eeqa}{\end{eqnarray}}
\newcommand{\bsub}{\begin{subequations}}
\newcommand{\esub}{\end{subequations}}

\newcommand{\rem}[1]{}
\newcommand{\refe}[1]{Eq.~(\ref{#1})}
\newcommand{\reff}[1]{Fig.~(\ref{#1})}

\newcommand{\EP}{\varepsilon_P}

\begin{document}
\title{
Signatures of the Current Blockade Instability in Suspended Carbon Nanotubes
}
\author{G. Micchi}
\author{R. Avriller}
\author{F. Pistolesi}
\affiliation{
Univ. Bordeaux, LOMA, UMR 5798,  Talence, France.\\
CNRS, LOMA, UMR 5798, F-33400 Talence, France.\\
}

\begin{abstract}
Transport measurements allow sensitive detection of nanomechanical motion of suspended carbon nanotubes. 
It has been predicted that when the electro-mechanical coupling
is sufficiently large a bistability with a current blockade appears.
Unambiguous observation of this transition by current measurements may be difficult.
Instead, we investigate the mechanical response of the system, namely
the displacement spectral function; the linear response to a driving; and the 
ring-down behavior.
We find that by increasing the electro-mechanical coupling 
the peak in the spectral function broadens and shifts at low frequencies 
while the oscillator dephasing time shortens.
These effects are maximum at the transition where non-linearities dominate the dynamics.
These strong signatures open the way to detect the blockade transition 
in devices currently studied by several groups. 
\end{abstract}

\date{\today}



\maketitle

\newcommand{\Eex}{E_{\rm ex}}
\newcommand{\com}[1]{{\tt #1}}

Recently enormous progress has been achieved in the detection of carbon nanotubes (CNT) bending modes by electronic transport measurements
\cite{sazonova_tunable_2004,
lassagne_ultrasensitive_2008,
steele_strong_2009,
lassagne_coupling_2009,
eichler_nonlinear_2011,
laird_high_2011,
chaste_nanomechanical_2012,
meerwaldt_probing_2012,
ganzhorn_dynamics_2012,
moser_nanotube_2014,
zhang_interplay_2014,
benyamini_real-space_2014,
schneider_observation_2014}.
Since nanotube oscillators have remarkable mechanical properties, devices with record mass \cite{chaste_nanomechanical_2012} and force \cite{moser_nanotube_2014} sensitivity have been realized.
Transport experiments allow information to be obtained on the mechanical mode by measuring different quantities.
The main ones are the oscillation amplitude in response 
to an external drive \cite{sazonova_tunable_2004}, the 
oscillator displacement spectral density, $S_{xx}(\omega)$ \cite{moser_nanotube_2014},
and, more recently, the ring-down time of the oscillator \cite{schneider_observation_2014}.

The recent experimental advances allow one to view
the behavior of such systems in the strong coupling limit
from a new perspective. 
Defining $F_0$ as the difference of electrostatic force acting on the nanotube when one electron is added to the suspended part and $k$ as the spring constant for its displacement, one can introduce a polaronic energy scale $\EP=F_0^2/k$.
For a classical system resonating at pulsation $\omega_0$ (much smaller than the bias voltage $eV$ or the temperature $T$) it has been predicted \cite{galperin_hysteresis_2005,mozyrsky_intermittent_2006,pistolesi_current_2007,pistolesi_self-consistent_2008} that, if $\EP \gg eV$ and $T$, the current in the device can be blocked and a bistability can appear
($e$ is the electron charge and we set both the Planck and Boltzmann constant to 1).
The energy $\EP$ can be estimated for actual experiments: For a CNT of 1 $\mu$m length, 1 nm radius, and suspended at a distance $d=500$ nm from a gate one finds $\omega_0/2\pi \approx$ 50 MHz, $F_0 = 10^{-14}$ N, $k=4 \cdot 10^{-4}$ N/m, and thus $\EP \approx 16$ mK.
One should thus work at very low temperatures $ \ll 16$ mK: 
This is probably why the current blockade transition has not yet been observed for mechanical bending modes. 
While a (Franck-Condon) blockade \cite{braig_vibrational_2003,koch_franck-condon_2005} has been observed \cite{leturcq_franckcondon_2009} for breathing modes in the regime of incoherent transport.
To increase $\EP$ one can reduce the distance of the CNT from the gate electrode or operate the system close to the Euler buckling instability \cite{weick_discontinuous_2010,weick_euler_2011}.
The energy $\EP$ scales quadratically with $d$, it is thus realistic 
to increase this energy up to the Kelvin range by reducing the distance  $d$ to 100 nm.
In any case, a clear observation of the transition will require temperatures of the order 100 mK.
At such low temperatures the typical tunnelling rate $\Gamma$ becomes larger than $T$, leading to coherent transport through the CNT.
In this regime the current blockade can take place only if $\EP$ is larger than a critical value
$\varepsilon_c$ of the order of $\Gamma$ \cite{galperin_hysteresis_2005,mozyrsky_intermittent_2006}.
Since one of the main experimental difficulties is to reach large values of $\EP$, the case $\EP \sim \Gamma$ 
is particularly interesting.
The transition could then be investigated at fixed and low $V$ and $T$ by varying $\EP$, that can be tuned with the gate voltage.
One drawback of this limit is the large width $\Gamma$ of the electronic level: 
The conductance dependence on $\EP$ is smooth on the scale $\Gamma$ failing to 
provide a clear indication of the transition.

In this Letter we show that, similar to critical phenomena, the transition can be better investigated by looking 
at the behavior of the phonon mode that becomes soft for $\EP=\varepsilon_c$.
We study $S_{xx}(\omega)$, the driving response function, and the dephasing and ring-down time 
as a function of the coupling constant $\EP$.
We find that all of these quantities have a very peculiar behavior at the transition.
The dynamics of the mechanical mode is dominated by non-linear terms 
leading to a separation of time scales that is maximal at the transition. 
The theory presented gives clear indication on how to 
unambiguously observe the transition using available methods of measurement.

{\em The model.} 
We consider a suspended CNT. 
We assume that a single electronic level is relevant for transport. 
We neglect the spin degrees of freedom.
The Hamiltonian reads:
\beq
	H=H_L+H_R+H_T+(\epsilon_0-x F_0) d^\dag d^{\phantom \dag}+{p^2\over 2m}+{k\over 2}x^2
	\,,
\eeq
where $d$ is the destruction operator for the electronic level on the dot, $x$ is the displacement of the relevant mechanical mode, $p$ the conjugated momentum, $m$ the mode effective mass, $k$ the spring constant (giving a pulsation $\omega_0=\sqrt{k/m}$) and $F_0$ the electrostatic force acting on the dot when an electron is added.
The first three terms describe the leads and their coupling: 
$H_\alpha=\sum_k (\epsilon_{\alpha k}-\mu_\alpha)  c_{\alpha k}^\dag c_{\alpha k}^{\phantom \dag}$, 
with $\alpha=L$ and $R$, for left and right lead, $\epsilon_{\alpha k}$ the electronic spectrum, $\mu_\alpha$ the chemical potential;
and $H_T=\sum_k t_{\alpha} c_{\alpha k}^\dag d^{\phantom \dag} + {\rm h.c.}$ the tunnelling Hamiltonian.
From these quantities one can define the single-level width $\Gamma_\alpha \equiv \pi t^2_\alpha \rho_\alpha$ with $\rho_\alpha$ the density of states and $\Gamma=\Gamma_L+\Gamma_R$. 

In the Born-Oppenheimer limit ($\Gamma_\alpha \gg \omega_0$) the displacement of the mechanical mode can be described by a Langevin equation:
\beq
	m\ddot x+A(x) \dot x+m\omega_0^2 x=F_e(x)+\xi(t) \, ,
\eeq
where the dissipation $A(x)$, the average force $F_e(x)=F_0\langle d^\dag d\rangle$, and the stochastic force $\xi(t)$ are due to the electrons tunneling through the quantum dot \cite{mozyrsky_intermittent_2006,pistolesi_current_2007}.
The explicit expressions for $A$, $F_e$, and $\langle \xi(t)\xi(t')\rangle=D(x) \delta(t-t')$ have been obtained in Ref.~\cite{mozyrsky_intermittent_2006}:
\beq
	F_e(x) = F_0
	\left[
	{1 \over 2} + {1 \over \pi} \sum_\alpha {\Gamma_\alpha \over \Gamma}
	\arctan {\mu_\alpha-\epsilon_0-F_0 x \over \Gamma} 
	\right]
		,
\eeq
$
	A(x)= {(F^2_0\Gamma/\pi)} \sum_\alpha 
	{\Gamma_\alpha/ [(\mu_\alpha-\epsilon_0-F_0 x)^2+\Gamma^2]^2}
$, and 
$
	D(x)= F^2_0\Gamma_L \Gamma_R/(\pi \Gamma^3)[h(\mu_L)-h(\mu_R)]
$,
where $h(\mu)=\arctan z+z/(z^2+1)$ with $z=(\mu-\epsilon_0-F_0 x)/\Gamma$.
In the same limit a Fokker-Planck equation for the probability $P(x,p,t)$ can be derived \cite{blanter_single-electron_2004,blanter_erratum:_2005}:
\beq
	\partial_t P={p\over m} \partial_x P-F \partial_p P + {A\over m} \partial_p (pP)+{D\over 2} \partial_p^2 P
	\,,
	\label{FP-eq}
\eeq
with $F(x)=F_e(x)-k x$.

\begin{figure}[tbp]
\includegraphics[width=.95\linewidth]{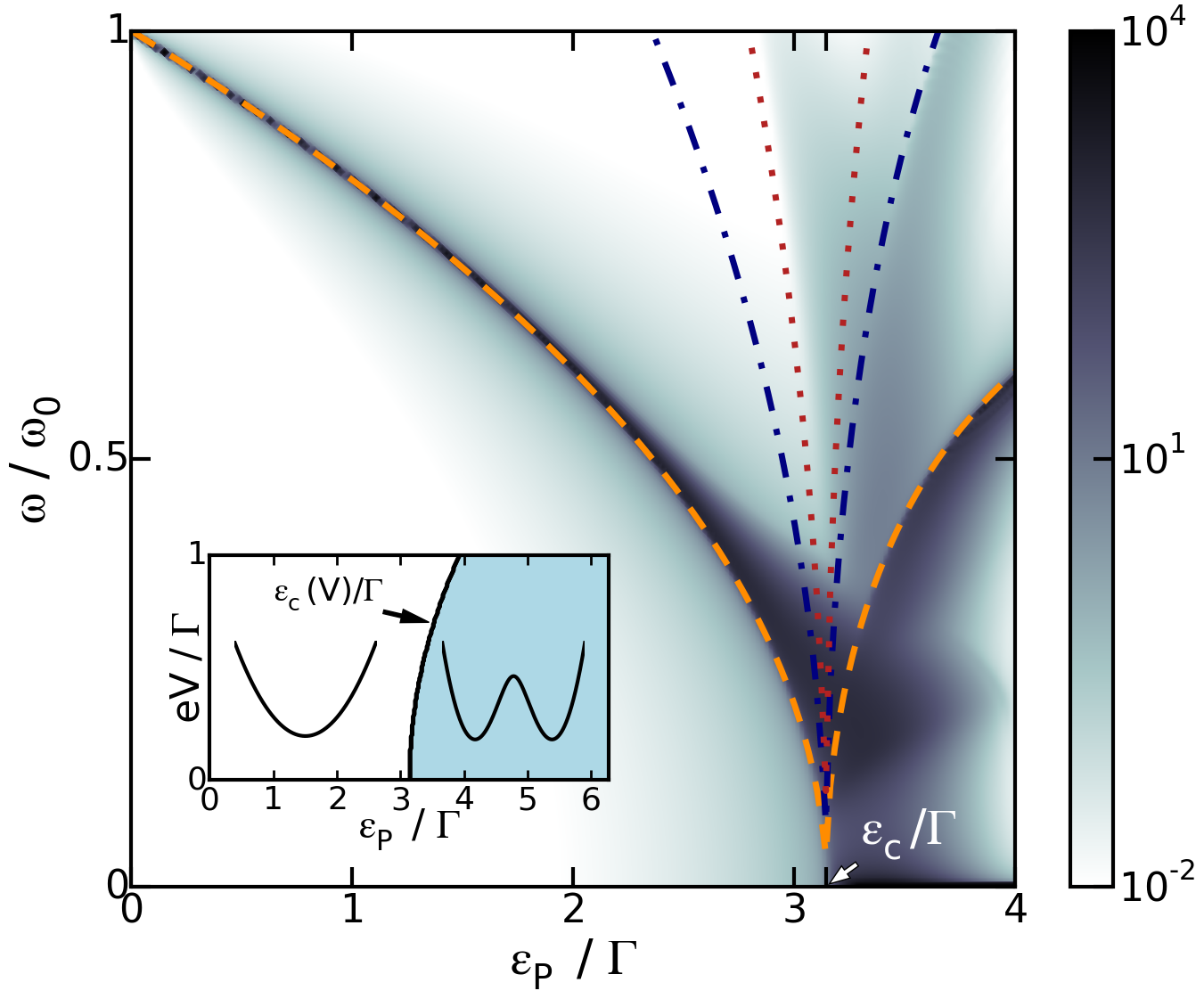}
  \caption{Density plot of $S_{xx}$ as a function of $\omega$ and $\EP$. 
The values of $\omega_m$ (orange dashed line), $2\omega_m$ (blue dot-dashed line),
and $3\omega_m$ (red dotted line) are shown.
The units of $S_{xx}$ are $x_{\rm zpm}^2/\omega_0=(m\omega_0^2)^{-1}$, 
where $x_{\rm zpm}=(m\omega_0)^{-1/2}$ is the zero-point motion displacement.
The symmetry of the potential implies that only odd harmonics are present for $\EP<\varepsilon_c$. 
Inset: phase diagram in the plane $eV$-$\EP$ for the stability of the effective potential.
}
\label{fig:spectrum}
\end{figure}

%
{\em Softening of the mechanical mode.}
We assume that the device is symmetric: $\Gamma_\alpha=\Gamma/2$.
The presence of the mechanical coupling modifies the electron-hole symmetry point for $\epsilon_0$ 
to the value $\epsilon_0=(\mu_L+\mu_R)/2+\EP/2$. 
We will always assume this value from this point on.
Defining $y=x-F_0/2k$, one determines
$F(y)=-ky+(F_0/2\pi)\sum_{a=\pm1}\arctan[(F_0 y+ a eV/2)/\Gamma] $ 
which depends only on the bias voltage $eV=\mu_L-\mu_R$ and is anti-symmetric in $y$.
The equilibrium positions are defined by the solutions of the equation $F(y)=0$.
The line $\EP=\varepsilon_c(V) \equiv  \pi  \Gamma [1+(eV/2\Gamma)^2]$ [for $eV/\Gamma<2/\sqrt 3$]
separates the monostable region from the bistable region (see inset of ~\reff{fig:spectrum}).

%
%
%
%
%

Let us now define $\omega_m$ at a stable point $y_\beta$ as $m \omega_m^2 = -(dF/dy)_{y_\beta}$.
It goes smoothly from $\omega_0$ to 0 for $\EP<\varepsilon_c$ with the analytic form $\omega^2_m/\omega^2_0=(\varepsilon_c-\EP)/\varepsilon_c$, while for $\EP \gtrsim \varepsilon_c$ it reads $2(\EP-\varepsilon_c)/\varepsilon_c$ [see~\reff{fig:spectrum} orange dashed lines].

The vanishing of $\omega_m$ suggests that its direct measurement 
should allow the detection of the transition with great accuracy.
As in phase transitions, this mode becomes soft, leading to a strong response at the transition. 
The dip in the gate voltage dependence of $\omega_m$ observed by four different groups     
\cite{steele_strong_2009,lassagne_coupling_2009,ganzhorn_dynamics_2012,benyamini_real-space_2014}
is the precursor of this softening.
Nevertheless, one should be cautious since the definition of $\omega_m$ 
only takes into account the first derivative of the force
at the minimum of the potential.
When this term vanishes the next order terms in $y$ become important 
and the response of the system can no longer be predicted simply by the value of $\omega_m$.
Therefore, in the following we calculate the typical measurable quantities 
and study their behavior when $\EP$ is swept through the transition.

%
{\em Fluctuation spectrum.}
We define the displacement fluctuation spectrum $S_{xx}(\omega)= \int e^{i\omega t}  dt \langle \tilde x(t) \tilde x(0)\rangle$, with $\tilde x(t)=x(t)-\langle x \rangle$.
This quantity has been measured recently in Ref.~\cite{moser_nanotube_2014}.
We can obtain $S_{xx}$ numerically from the Fokker-Planck description following the method used in Ref.~\cite{pistolesi_self-consistent_2008}.
Writing \refe{FP-eq} as  $\partial_t P={\cal L}_0 P$ the spectrum takes the form:
\beq
	S_{xx}(\omega)= -2 {\rm Tr} 
	\left[ 
	\hat {\tilde x} {{\cal L}_0 \over \omega^2 +{\cal L}_0^2} \hat {\tilde x} P_{\rm st}
	\right]
	\label{eq:Sxx}
\eeq
where $P_{\rm st}$ is the stationary solution of the problem, satisfying both ${\cal L}_0 P_{\rm st}=0$ 
and the normalization condition ${\rm Tr} P_{\rm st}=1$.
The operator $\hat {\tilde x}$ is defined as $\hat {\tilde x} P \equiv \tilde x P(x,p)$.

Let's begin by discussing the stationary solution of \refe{FP-eq}: $P_{\rm st}$.
In agreement with similar models \cite{armour_classical_2004,weick_euler_2011} we find that for 
sufficiently small $eV$, even if the system is out of equilibrium, the stationary distribution 
function takes the simple Gibbs form $P_{\rm st}(x,p)={\cal N} \exp\{-E(x,p)/T_{\rm eff} \}$, 
with  $T_{\rm eff}=eV/4$, ${\cal N}$ a normalization factor, $E(x,p)=p^2/2m+U(x)$, 
$U(x) = -dF/dx$, and $U(x)=0$ at its minimum.
This result is due to the smooth dependence on $x$ of both $D(x)$ and $A(x)$ on the scale of the spread of 
the probability distribution $P(x,p)$ for $eV\ll \Gamma$.

\begin{figure}[tbp]
\includegraphics[width=\linewidth]{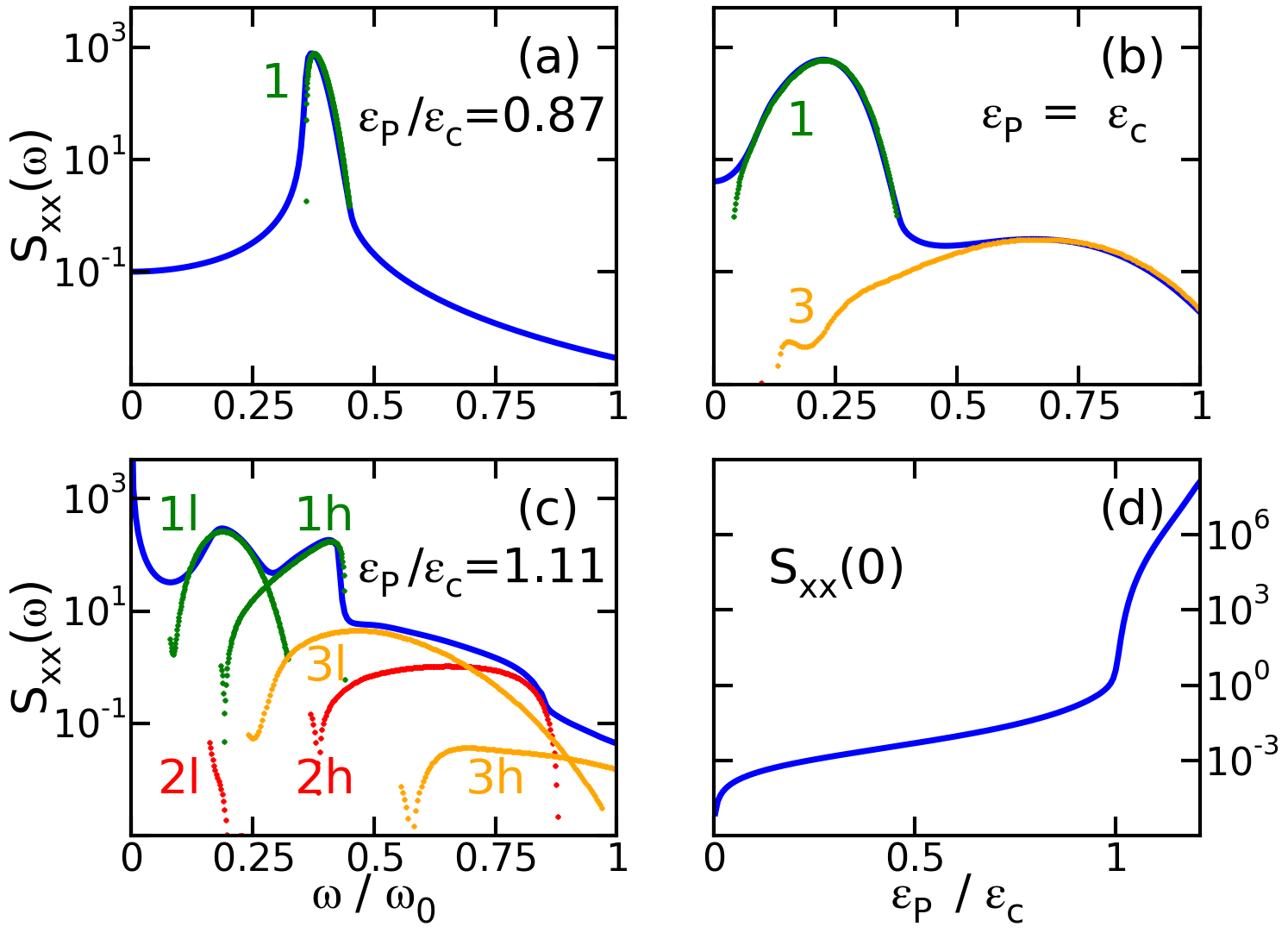}
\caption{Comparison of the full numerical solution of the Fokker-Planck equation for $S_{xx}$  (blue solid lines) with 
the one obtained with \refe{SxxAna} (dots) for $\EP/\varepsilon_c=0.87$, 1, 1.11 (a, b, and c  panel, respectively) in units of $(m\omega_0^2)^{-1}$.
The numbers label the order of the harmonic, while the letters h and l in the c panel 
indicate the high- and low-frequency contributions.
Panel d: $S_{xx}(0)$ as a function of $\EP/\Gamma$ indicating the onset of the telegraph noise at the transition.
}
\label{fig:SxxAna}
\end{figure}

We come now to the displacement spectrum obtained from \refe{eq:Sxx} that we show 
for $eV/\Gamma = 5 \cdot 10^{-3}$  in Fig.~\ref{fig:spectrum} and \ref{fig:SxxAna}.
As anticipated, as coupling increases the resonance broadens and shifts at low frequency.
More surprisingly, the peak position shows a minimum at a small but finite value of $\omega$ 
at the transition with a maximal broadening (cf. Fig.~\ref{fig:SxxAna}-b). 
The position of the minimum and its width depend on the bias voltage.
A strong telegraph noise appears for $\EP \gtrsim \varepsilon_c$ (dark region at 
$\omega\rightarrow 0$), signaling the hopping of the systems between the two minima in the potential \cite{mozyrsky_intermittent_2006,pistolesi_self-consistent_2008}. 
This is also a strong indication of the transition [see \reff{fig:SxxAna}-d].
In the bistable phase ($\EP>\varepsilon_c$) a double peak in the spectral function is visible.

In order to understand this behavior we take advantage of the 
separation of time scales of the problem.
The damping $A(x)$ and the fluctuations $D(x)$ are both generated by the non-equilibrium 
electronic transport and, by hypothesis, are parametrically smaller (as $\omega_0/\Gamma$) 
than the Hamiltonian terms in \refe{FP-eq}. 
This implies that the system performs many oscillations 
on the closed trajectory in the phase space that satisfies $E(x(t),p(t))=E$, 
before drifting to a nearby trajectory 
on the slow time scale  $\gamma_E^{-1}$, where $\gamma_E=\int dx P_{\rm st}(x,p) A(x)/m$ is the average 
dissipation coefficient \cite{blanter_single-electron_2004,pistolesi_current_2007}.
For each energy $E$ one can then calculate the pulsation of the 
closed trajectory $\omega(E)=1/\left[2 \pi (m/2)^{1/2} \oint [E-U(x)]^{-1/2}dx\right] $.
The non-linearities present in $U(x)$ induce dispersion in $\omega(E)$.
Then, from our definition, $\omega_m$ indicates only $\omega(0)$.
Energies up to $eV$ are populated as an effect of the stochastic fluctuations.
They all contribute to the fluctuation spectrum leading to an inhomogeneous broadening
that spans the frequencies between $\omega_m$ and $\omega(eV)$. 

In order to provide a quantitative verification of this interpretation we calculate the 
spectrum by neglecting the effect of the dissipation and considering only the interference 
of the different trajectories populated according to $P_{\rm st}$ \cite{dykman_time_1980,dykman_spectral_1985}.
This gives 
$S_{xx}(t)=\int dx_0 dy_0 P_{\rm st}(x_0,p_0) \tilde x(t) \tilde x(0)$, 
where $x(t)$ satisfies the equation of motion $m\ddot x=F(x)$, with the initial conditions 
$x(0)=x_0$, and $\dot x(0)=p_0/m$.
In terms of the Fourier coefficients of fixed energy periodic trajectories
[$\tilde x_E(t)=\sum_n e^{i n \omega(E)t} x_n(E)$] the spectrum takes the form:
\beq
	S_{xx}(\omega)=\int_0^\infty {\cal P}(E) dE \sum_n 2 \pi \delta(\omega-n \omega(E)) x_n^2(E)
	\label{SxxAna}
	\,,
\eeq
with ${\cal P}(E)={\cal N} e^{-E/T_{\rm eff} }2\pi/\omega(E)$.

Two limits can be analyzed \footnote{See supplemental materials Sec. I}:
When the quartic term is much smaller than the quadratic one, 
the full width at half height of the resonance is 
$\Delta \omega \approx 1.5 \omega_0 (eV/\varepsilon_c)(\EP/\varepsilon_c)^2(1-\EP/\varepsilon_c)^{-3/2}$
with a small positive shift of the maximum from $\omega_m$ of the same order.
In the opposite limit of vanishing harmonic term ($\EP=\varepsilon_c$) 
the potential can be approximated as quartic. 
In this case $\Delta \omega/\omega_0=0.50 (eV/\Gamma)^{1/4}$ with a peak position $\omega_M$
at $0.85 \omega_0 (eV/\Gamma)^{1/4}$.
This voltage dependence can be related to the dispersion of 
$\omega(E)$ that vanishes as $E^{1/4}$ for $\EP=\varepsilon_c$.
Remarkably, at criticality the Q-factor of the oscillator defined as $\omega_M/\Delta \omega$
takes the {\em universal value} 1.71, independently of $V$ or $\Gamma$.
The crossover between the two regimes takes place for 
$1-\EP/\varepsilon_c \approx 1.71 (eV/\varepsilon_c)^{1/2}$, thus 
for the values considered in Fig.~\ref{fig:spectrum} the quartic region 
is restricted to $1-\EP/\varepsilon_c < 0.03$.
Finally, the double peak of the spectral function for $\EP > \varepsilon_c$ can be explained by 
the contributions of the low-energy high-frequency trajectories around each single minimum,
and those, at higher energy and lower frequency, revolving around both minima. 
The comparison with the numerical result presented in \reff{fig:SxxAna} shows 
a very good agreement. 
%
%
%
%

%
{\em Driving.}
Let us consider the other main tool used to detect mechanical motion: The response to a driving force of frequency $\omega_D$. 
We can find the linear response of the system by letting $F(x) \rightarrow F(x)+F_D \cos(\omega_D t)$ in \refe{FP-eq}. 
The evolution operator becomes $	{\cal L}(t)={\cal L}_0+ 2 {\cal L}_D \cos(\omega_D t)$, with ${\cal L}_D=-F_D \partial_p/2$.
After a transient time the solution can be written as a Fourier series $P(t)=\sum_n e^{i n \omega_D t} P_n$ where each Fourier component can in turn be expanded as a power series of the driving parameter $F_D$: $P_n=\sum_{k=0}^\infty P_{n,k}$, with $P_{n,k}$ of order $F_D^k$.
This leads to the equation for each component $(i n\omega_D -{\cal L}_0) P_{n,0}=0$ and 
\beq
(i n \omega_D -{\cal L}_0)P_{n,k+1} = {\cal L}_D (P_{n+1,k}+P_{n-1,k}) ,
\label{PnkEq}
\eeq
with the condition ${\rm Tr} P(t)=1$ . \refe{PnkEq} can be solved by recursion.
The time dependence of the displacement then reads $\tilde x(t)\equiv {\rm Tr} \left[\hat {\tilde x} P(t)\right]=F_D \chi(\omega_D) e^{i \omega_D t}+{\rm c.c.}$,
where 
\beq
\chi(\omega)=  {\rm Tr} \left[ \hat{ \tilde x} P_1(t) \right] = {\rm Tr} \left[ \hat{ \tilde x} (i\omega-{\cal L}_0)^{-1} \partial_p  P_{\rm st} \right]
	\,. 
		\label{eq:chi}
\eeq 

Naturally, the relation between $\chi(\omega)$ and $S_{xx}(\omega)$ comes into question.
If $P_{\rm st}$ has a Gibbs form then 
$F_D {\cal L}_0 \hat {\tilde x} P_{\rm st} = -2 T_{\rm eff} {\cal L}_D P_{\rm st}$.
This leads to a fluctuation-dissipation relation: 
\beq
	{\rm Im}\left[ \chi(\omega)\right]={\omega \over 2  T_{eff} } S_{xx}(\omega)
		\,.
		\label{eq:FD}
\eeq
Thus, for $eV \ll \Gamma$, $\chi$ and $S_{xx}$ give access to the same information 
in two independent ways.
For larger voltages expression (\ref{eq:chi}) always holds while \refe{eq:FD} will be violated.

%
{\em Ring-down behavior.}
Finally, let us consider the response for time  $t > 0 $ of the oscillator when the coherent drive 
is switched off at $t=0$. 
A damped harmonic oscillator relaxes exponentially on a time scale (the ring-down time) 
given by the same dissipation coefficient that also determines the width of the resonance 
of the response function.
For nano-mechanical oscillators it has recently been shown \cite{schneider_observation_2014} 
that this may not be the case. 
Non-linearities induce frequency noise, which in turn is responsible for 
phase fluctuations of $\tilde x(t)$.
The average over many realizations of $\tilde x(t)$ decays then on the time scale $\gamma_\varphi^{-1}$,
the value for which phase fluctuations become the order of $2\pi$.
Since the energy is insensitive to the phase, its average decays on the same time scale 
as the single realization $\gamma_E^{-1}$.
%

%
%
%
\begin{figure}[tbp]
\includegraphics[width=\linewidth]{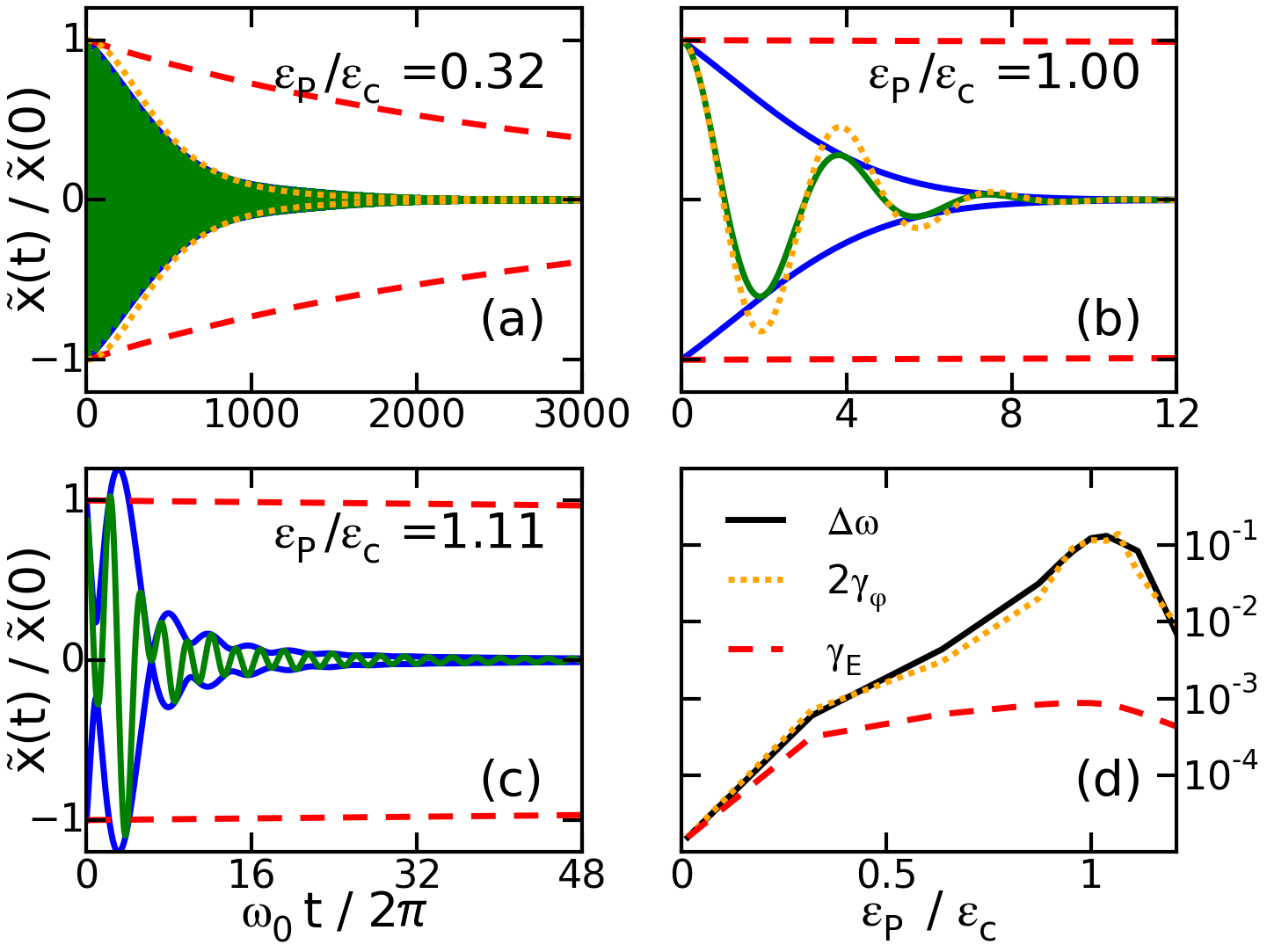}
\caption{
Time dependence of $\langle\tilde x(t)\rangle$ (green solid line) 
and its envelope (blue solid line)
for $\EP/\varepsilon_c=0.32$, 1, and 1.11 
(a, b, and c panel, respectively). 
The exponential decay on the scale $2/\gamma_E$ is shown red dashed.
In panel a and b the orange dotted line gives the result of 
the analytical expressions discussed in the text. 
Panel d: comparison of the  $\EP$-dependence of 
$\Delta \omega$, $2\gamma_\varphi$, and $\gamma_E$.
}
\label{fig:ring}
\end{figure}
%
%

With respect to our problem, we use the solution of the Fokker-Planck equation with driving 
[\refe{PnkEq}] as the initial condition and then find $\tilde x(t)$ and $E(t)$ 
from the evolution of the probability with ${\cal L}_0$.
The result as a Laplace transform reads:
\beq
	\langle\tilde x(z)\rangle ={\rm Tr} \left[ \hat {\tilde x} (z-{\cal L}_0)^{-1} P(t=0) \right] \, .
	\label{xofz}
\eeq
In a similar way we can calculate also the Laplace transform of the evolution of the 
total energy by letting $\hat {\tilde x} \to E(\hat x, \hat p)$ in \refe{xofz}.
One can then obtain the time dependence by numerically implementing the Cauchy 
theorem $	\langle\tilde x(t)\rangle = \oint_C \langle \tilde x(z)\rangle  e^{-z t}dz/(2\pi i)$, where $C$ is a 
contour that encloses 
the poles of $	\langle\tilde x(z)\rangle $ for ${\rm Re} z < 0$. 

We find that the energy exponentially decays on the scale $\gamma_E^{-1}$, even at the transition.
On the other hand, as shown in \reff{fig:ring}, 
$\langle\tilde x(t)\rangle$  decays on a much shorter scale that we define $\gamma_\varphi^{-1}$.
\reff{fig:ring}-d shows the $\EP$ dependence of $\Delta \omega$, $2\gamma_\varphi$, 
and $\gamma_E$.
The width $\Delta \omega$, obtained from the form of $S_{xx}$, coincides within the numerical 
accuracy with $2\gamma_\varphi$, proving that frequency noise is the responsible of the 
faster decay of $\langle\tilde x(t)\rangle$.
Both present a pronounced maximum at $\EP=\varepsilon_c$, indicating the transition.
Using the approach presented for the analytical calculation of $S_{xx}(\omega)$ we find 
that $\langle\tilde x(t)\rangle$ decays as $1/(1+t^2\gamma_\varphi^2)^2$, where 
$\gamma_\varphi=0.41\Delta \omega $ \footnote{see supplemental materials Sec. II}.
Similarly for $\EP=\varepsilon_c$ the decay scale is proportional to $\Delta\omega$ with 
the analytical form  given by the dotted line of  \reff{fig:ring}.

%
{\em Conclusions.}
We found that the study of the mechanical properties of the suspended carbon nanotube 
open new perspectives for the observation of the current-blockade transition occurring 
at low temperaure and voltage for $\EP = \pi \Gamma$. 
Indeed, for that value the quadratic part of the effective potential vanishes, leading to 
strong frequency and phase fluctuations that remarkably modify the typically
measured response functions  ($S_{xx}$, $\chi$, and ring-down behavior).
The $Q$-factor of the resonator takes the minimum and universal value 1.71 
at criticality, where the separation of time scales is also maximal. 
These results can lead to the observation of the current-blockade transition
by mechanical measurements in devices currently investigated by several experimental groups. 

{\em Acknowledgements.} We acknowledge support from ANR-10-BLANC-0404 QNM. 

\bibliography{Full}

\end{document}